# Orbital angular momentum divider of light

Hailong Zhou,[1] Jianji Dong,[1,*] Shimao Li,[2] Xinlun Cai,[2] SiYuan Yu[2] and Xinliang Zhang[1]

[1]*Wuhan National Laboratory for Optoelectronics, School of Optical and Electronic Information, Huazhong University of Science and Technology, Wuhan, China, 430074*

[2]*State Key Laboratory of Optoelectronic Materials and Technologies and School of Physics and Engineering, Sun Yatsen University, Guangzhou 510275, China.*

*Corresponding author: jjdong@hust.edu.cn*

**Abstract:** Manipulation of orbital angular momentum (OAM) of light is essential in OAM-based optical systems. Especially, OAM divider, which can convert the incoming OAM mode into one or several new smaller modes in proportion at different spatial paths, is very useful in OAM-based optical networks. However, this useful tool was never reported yet. For the first time, we put forward a passive OAM divider based on coordinate transformation. The device consists of a Cartesian to log-polar coordinate converter and an inverse converter. The first converter converts the OAM light into a rectangular-shaped plane light with a transverse phase gradient. And the second converter converts the plane light into multiple diffracted light. The OAM of zeroth-order diffracted light is the product of the input OAM and the scaling parameter. The residual light is output from other diffracted orders. Furthermore, we extend the scheme to realize equal N-dividing of OAM and arbitrary dividing of OAM. The ability of dividing OAM shows huge potential for OAM-based classical and quantum information processing.

PACS number(s): 42.50.Tx, 42.40.Jv, 42.30.Kq.

## I. INTRODUCTION

Light carrying orbital angular momentum (OAM) has recently attracted great interest due to their wide applications, such as micromanipulation [1,2], probing the angular velocity of spinning microparticles or objects [3,4], quantum information [5,6] and optical communication [7,8]. Here, the electromagnetic field of an OAM light is identified by a helical phase structure expressed as $\exp(il\theta)$, where $\theta$ is the angular coordinate, and the integer $l$ is the azimuthal index, indicating the topological charge (TC) of the OAM light [9]. The OAM light could be generated by various methods, such as spiral phase plate, fork grating, metasurface and q-plate [10-13]. These methods could also be applied to shift an OAM mode to another one, acting as OAM shifters. There were also many other functional devices highly developed, such as OAM mode filter, OAM add/drop multiplexer, OAM switch [14-16] and OAM sorter [17-19]. These functional devices make the use of OAM more flexible and easier. Even so, some vital functions for OAM-based applications are still absent and need to be explored.

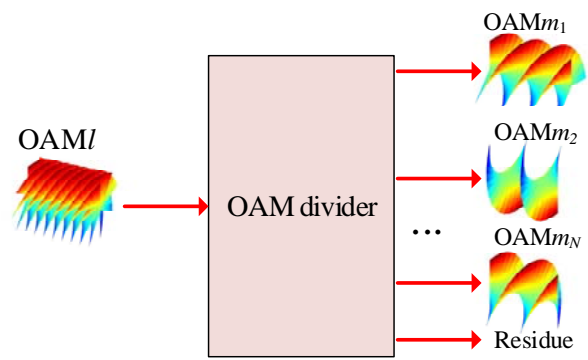

FIG. 1. (Color online) Sketch of OAM divider. The OAM divider can convert the OAM mode into several new OAM modes in proportion at different spatial paths.

In OAM-based classical and quantum systems, it is desirable to convert the input OAM mode into one or several new OAM modes as new channel labels or quantum states for router, switching, or transmission. As shown in Fig. 1, there is a conceptual device which can convert the OAM mode into several new OAM modes in proportion at different spatial paths, named as OAM divider. These new OAM modes can act as new channels for communication. When an OAM mode with TC = $l$ (labeled as $\text{OAM}l$) is incident on the device, the OAM divider will convert the OAM mode into several different OAM modes ($\text{OAM}m_1$, $\text{OAM}m_2$, ... , $\text{OAM}m_N$), and the residual light will be output from other ports. The condition needs to be met by

$$|l| \geq \sum_{n=1}^{N} |m_n|. \quad (1)$$

The equality operator is satisfied only when there is no residual light output. Although the dividing of OAM is an important function to manipulate the mode or the quantum state in OAM-based system, OAM divider has been rarely reported and achieved yet.

Recently, a very successful method to sort OAM modes by employing a Cartesian to log-polar transformation has been demonstrated [17,20,21]. By this transformation, the OAM mode was mapped to a rectangular-shaped plane light with a transverse phase gradient. And in turn, the tilted planar wavefront could be converted into an azimuthal phase profile by reverse input [22]. Combining these two characteristics, it is possible to reshape the OAM mode by cascading the two coordinate transformation devices.

In this article, we report a passive scheme to divide the OAM of light proportionally for the first time. The device consists of a Cartesian to log-polar coordinate converter and an inverse converter. The first converter consisting of two diffractive optical elements performs a Cartesian to log-polar coordinate transformation, aiming to convert the OAM light into a rectangular-shaped plane light with a transverse phase gradient. And the second converter performs an inverse coordinate transformation with an adjustable scaling parameter, aiming to convert the plane light into multiple diffracted light. The OAM of zeroth-order diffracted light is the product of the input OAM and the scaling parameter. The residual light is separated from other diffracted orders. Furthermore, we extend the scheme to implement equal *N*-dividing of OAM and arbitrary dividing of OAM. The dividing of OAM shows huge potential for OAM-based classical and quantum information processing.

## II. PRINCIPLE

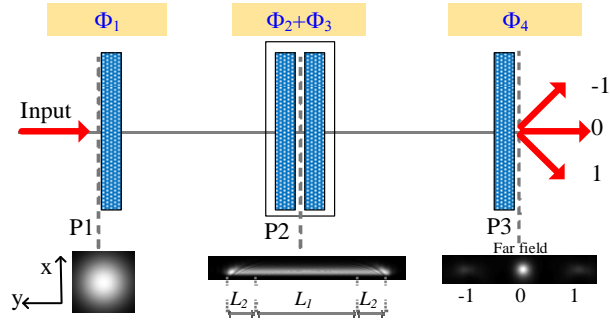

FIG. 2. (Color online) Schematic diagram for dividing of OAM. The first two optical elements convert the OAM light into a linear light with a transverse phase gradient. The last two convert the linear light into multiple light.

The schematic diagram for dividing of OAM is shown in Fig. 2. The first two diffractive optical elements perform a Cartesian to log-polar coordinate transformation from Plane P1 to P2. When an OAM light is incident on the device, the annular-shaped light with a helical phase structure will be converted into a rectangular-shaped plane light with a transverse phase gradient. To achieve the mapping of a position $(x, y)$ in Plane P1 to a position $(u, v)$ in Plane P2, where $u = -a \ln(\sqrt{x^2 + y^2}/b)$ and $v = a \arctan(y/x)$, the phase profiles of the two optical elements are given by [17,20,21]

$$\Phi_1 = \Phi_{T1}(x,y,a,b) + \Phi_L(x,y)$$

$$= \frac{2\pi a}{\lambda f}\left[ y\arctan(\frac{y}{x}) - x\ln(\frac{\sqrt{x^2+y^2}}{b}) + x - \underbrace{\frac{1}{a}\left(\frac{x^2+y^2}{2}\right)}_{lens\ term} \right], (2)$$

$$\Phi_2 = \Phi_{C1}(u,v,a,b) + \Phi_L(u,v)$$

$$= \frac{-2\pi ab}{\lambda f}\left[ \exp(-\frac{u}{a})\cos(\frac{v}{a}) + \underbrace{\frac{1}{ab}\left(\frac{u^2+v^2}{2}\right)}_{lens\ term} \right], (3)$$

where $\lambda$ is the wavelength of the incoming beam, and $f$ is the distance between the two elements. $\Phi_{T1}$ is the phase profile for implementing coordinate transformation, $\Phi_{C1}$ is the required phase correction and $\Phi_L$ is a lens term for lensless system. The parameter $a$ scales the transformed image and $a = d/2\pi$, where $d$ is the length of the rectangular-shaped light. $b$ translates the rectangular-shaped light in the $u$ direction and can be chosen independently of $a$. The last two optical elements constitute an inverse coordinate converter, the phase profiles are given by

$$\Phi_3 = \Phi_{T2}(u,v,a,b) + \Phi_L(u,v)$$

$$= \frac{-2\pi\gamma ab}{\lambda f}\left[ \exp(-\frac{u}{\gamma a})\cos(\frac{v}{\gamma a}) + \underbrace{\frac{1}{\gamma ab}\left(\frac{u^2+v^2}{2}\right)}_{lens\ term} \right], (4)$$

$$\Phi_4 = \Phi_{C2}(x,y,a,b) + \Phi_L(x,y)$$

$$= \frac{2\pi\gamma a}{\lambda f}\left[ y\arctan(\frac{y}{x}) - x\ln(\frac{\sqrt{x^2+y^2}}{b}) + x - \underbrace{\frac{1}{\gamma a}\left(\frac{x^2+y^2}{2}\right)}_{lens\ term} \right], (5)$$

where $\gamma$ is an adjustable scaling parameter and $|\gamma| \leq 1$. $\Phi_{T2}$ is used to implement an inverted coordinate transformation and $\Phi_{C2}$ corrects the phase. The inverse coordinate converter can convert the middle part of rectangular-shaped light into an annular-shaped light and the light will be output from the zeroth-order. The residual light is separated from other diffracted orders. Here, the length of the middle part is $L_1 = |\gamma| d$. Figure 2 presents the pattern conversion for $\gamma = 0.5$ when a Gaussian mode (OAM0) is incident from Plane P1. We can see that the Gaussian mode is converted into a rectangular-shaped light in Plane P2 and then is converted into a spot in the far field of Plane P3. There is also two residual light output from the $\pm 1st$-order.

Figure 3(a) presents the detailed transformation process for an OAM light (taking OAM4 as an example). The parameters are set as $f = 210mm$, $a = 1.5mm$ $b = 1mm$, $\lambda = 1.5 \mu m$, and the beam waists of input modes are about 1~2 *mm*. The flatbed grayscale images refer to the phase patterns used in the simulations and the plane images show the intensity patterns of light in various planes. For the sake of analysis, all the four phase patterns are divided into a same lens term and another special term. For a doughnut-shaped light, the annulus is unwrapped and gradually straightened after the first element. It is fully straightened into a rectangular-shaped light with an additional phase when arriving at the second element. Subsequently the additional phase is corrected by $\Phi_{C1}$ and the rectangular-shaped light become a plane wave. $\Phi_{T2}$ is used to implement an inverted coordinate transformation. If $\Phi_{T2} = \Phi_{C1}$, i.e., the cosine-like period of $\Phi_{T2}$ (marked by red rectangle) exactly overlaps with the rectangular-shaped light (marked by yellow rectangle), the output light and input light will be same. And when we shorten the period of $\Phi_{T2}$, only the light in the cosine-like period is wrapped and evolved into an doughnut-shaped light. The light will output from the zeroth order and the OAM will be reduced proportionally. The residual light in other two cosine-like periods will output from the $\pm 1st$-order. Figure 3(b) shows the simulated field distributions in various planes for $\gamma = 0.5$ when OAM4, OAM6, OAM8 are incident respectively. One can see that the zeroth-order diffracted light is an OAM mode with TC equal to $l/2$ when an OAM$l$ is incident from Plane P1, and the residual light is output from other diffracted orders. In addition, there is almost no crosstalk observed in the output light for the divider because the angle spacing of output light is quite large.

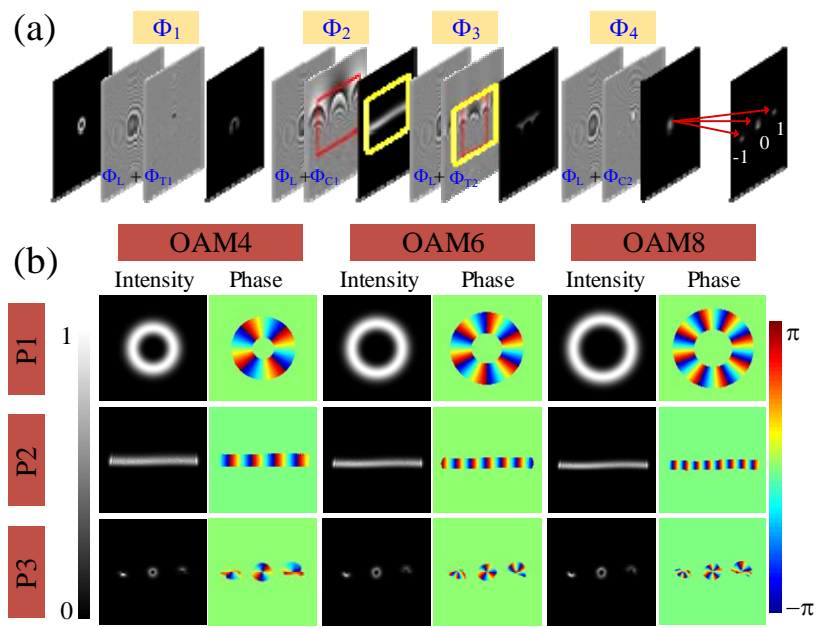

FIG. 3. (Color online) (a) Transformation process of OAM divider. (b) Optical field distributions in various planes (P1, P2, P3) when inputting OAM4, OAM6, OAM8 respectively.

In a real deployment, the second and third elements can be combined to one element, since they are both in Plane P2. So the device consists of three diffractive optical elements actually.

## III. EXPERIMENT

A proof-of-concept experiment is designed to demonstrate the scheme, as shown in Fig. 4. The light emitted from a laser (wavelength at 1550 nm) is expanded with a collimator, then a polarizer is used to match the operating polarization of spatial light modulators (SLMs). The first SLM (SLM1) is used to generate the input OAM light. The OAM light is diffracted from the first-order by adding a grating in SLM1 and then it is selected with a pinhole. The followed three SLMs (SLM2, SLM3 and SLM4) constitute the OAM divider shown in Fig. 2. Afterwards, a lens is employed to separate the multistage diffracted light and the final output light is collected by a charge-coupled device (CCD). In the experiment, the parameters are set as the same with the simulations.

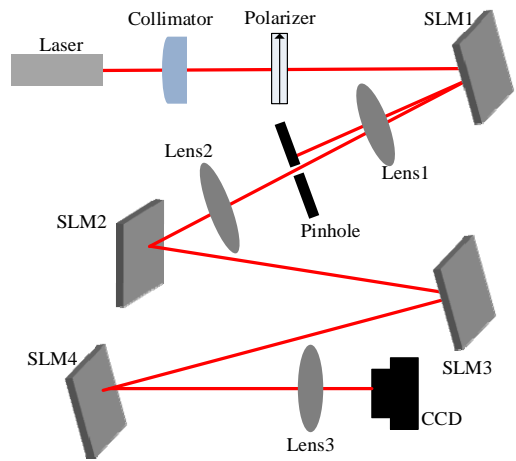

**FIG. 4. (Color online) Experimental setup. SLM1 is used to generate the input OAM light. SLM2, SLM3 and SLM4 constitute the OAM divider.**

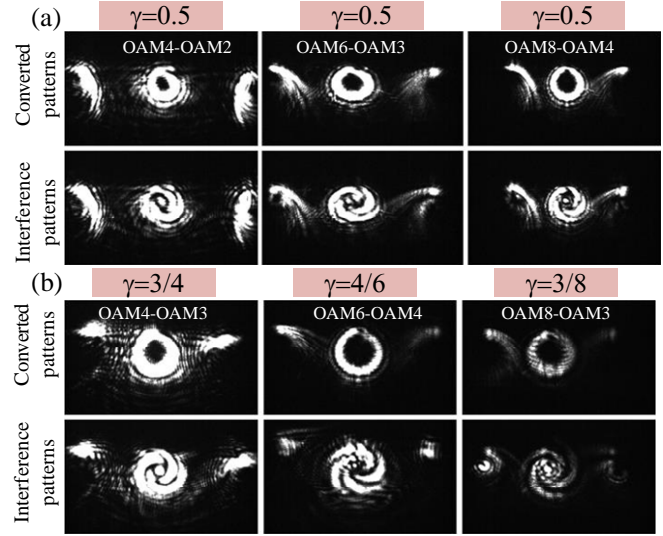

FIG. 5. (Color online) Experimental results. (a) The first row shows the output light for $\gamma = 0.5$ when OAM4, OAM6, OAM8 are incident respectively. The second row shows the interference patterns. (b) Results for different scaling parameters.

Firstly, we experimentally verify the simulated results in Fig. 3(b). As shown in Fig. 5(a), the first row shows the output light collected by CCD for $\gamma = 0.5$ when OAM4, OAM6, OAM8 are incident respectively. One can see that there are three diffracted light collected. The zeroth-order diffracted light is a fully annular light and the $\pm 1st$-order light reveals irregular shapes. In order to measure the TC of output light, we introduce a referenced light (OAM0) by SLM1 to generate interference patterns. The interference patterns are presented in the second row. As expected, the interference patterns indicate that the TCs of zeroth-order diffracted light are all the half of input ones. We also verify other proportional dividing of OAM. Figure 5(b) shows the results when the scaling parameters are 3/4, 4/6, 3/8 respectively. In order to get integral TCs, the input OAM modes are adjusted accordingly. The second row of Fig. 5(b) shows the interference patterns, verifying the correctness of the OAM divider. The overall efficiency of the divider is about 20% and the losses mainly come from the low diffraction efficiencies of SLMs (~60%). The overall efficiency can be further raised to near 70%, if we replace the SLMs with refractive elements [21]. In addition, some overlaps can be observed from the experimental results. It is caused by that the observing plane slightly deviates from the focal plane, because the size of three output light in focal plane is too small to observe clearly for our CCD.

## IV. DISCUSSION

In the above scheme, we successfully demonstrate the proportional dividing of OAM, but only the zeroth-order light is converted into an OAM mode, and other orders of diffracted light are discarded. In fact, all orders of the diffracted light can be converted into OAM modes via proper design. For the inverse coordinate converter, $\Phi_4$ is the required phase correction which is related to the scaling parameter. When we divide the rectangular-shaped plane light in Plane P2 into *N* equal segments and perform an inverse coordinate transformation on every segment independently, all the output images of *N* segments in Plane P3 will be the same but with different output directions. So all the phase corrections of *N* segments are the same and can be completed by $\Phi_4$. In the case, the phase profiles are given by

$$\Phi_3 = \Phi_L(u,v) + \sum_{n=1}^{N} \Phi_{C1}(u, v - v_n, \sigma a/N, b)(v \in L_n), \quad (6)$$

$$\Phi_4 = \Phi_{T1}(x, y, \sigma a/N, b) + \Phi_L(x, y). \quad (7)$$

Here, $v_n$ is the central position of *n*th segment, $L_n$ indicates the region of *n*th segment, and $\sigma = \pm 1$, representing the sign of scaling conversion. Figure 6 presents simulated examples when an OAM6 is incident. To convert the rectangular-shaped light into *N* annular light, *N* cosine-like phase patterns are required to reshape the *N* segments of light independently. As shown in Fig. 6(a), the period of $\Phi_{T2}$ is dropped to one *N*-th of the length of rectangular-shaped light and the central position of every period is adjusted accordingly to cover all the light. The first row of Fig. 6(b) shows the intensity distribution and phase distribution in the far field when dividing the input OAM6 into two equal modes, we can see two OAM3 are generated in two directions. Furthermore, three OAM2 are generated in three directions when dividing the input OAM mode into three equal modes.

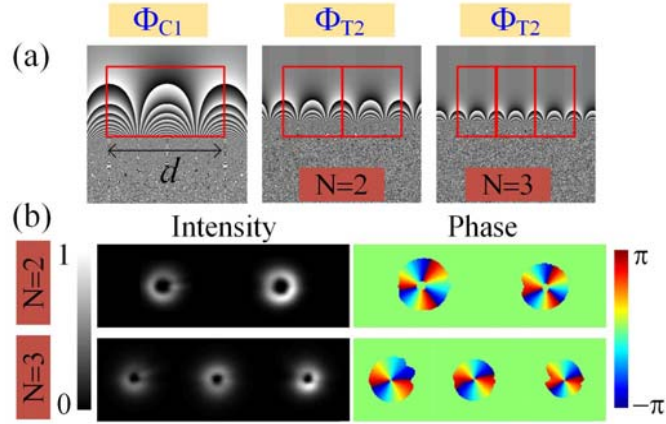

FIG. 6. (Color online) Equal *N*-dividing of OAM. (a) The needed phase patterns and (b) simulated results when an OAM6 is incident.

The lengths of *N* segments can also be different if the images are separated in Plane P3. According to the imaging theory of lens, the image in the Fourier plane can be shifted by adding an additional gradient phase on the input plane. As shown in Fig. 7(a), we add a corresponding gradient phase on every segment $L_n$ in Plane P2 to make the corresponding output image completely separated in Plane P3 and then the phase can be corrected independently in Region $R_n$ of Plane P3. Here, $R_n$ is the assigned region in Plane P3 for the output image of *n*th segment. In the case, the phase profiles are revised by

$$\Phi_3 = \Phi_L(u,v) + \sum_{n=1}^{N}\left[\Phi_{C1}(u, v-v_n, a_n, b) + \frac{2\pi}{\lambda f}(v_{nx}u + v_{ny}v)\right](v \in L_n), (8)$$

$$\Phi_4 = \Phi_L(x,y) + \sum_{n=1}^{N}\left[\Phi_{T1}(x-v_{nx}, y-v_{ny}, a_n, b) + \frac{2\pi}{\lambda f}v_n y\right]((x,y) \in R_n). (9)$$

Here, $(v_{nx}, v_{ny})$ is the desired center coordinate of the output image of *n*th segment in Plane P3 and $a_n = \pm aL_n/d$. The gradient phase term in Eq. (9) is used to correct the direction of the *n*th output image. According to Eqs. (8, 9), the condition is satisfied: $|TC_{in}| = \sum_{n=1}^{N}|TC_n|$, where $TC_{in}$ is the TC of input light and $TC_n$ is the TC of *n*th output image. Figure 7(b) shows two simulated examples when an OAM8 is incident. The first row shows the output patterns in Plane P3 in the case where the OAM8 is divided into an OAM4 and two OAM-2. The second row shows the output patterns in Plane P3 in the case where the OAM8 is divided an OAM5 and an OAM-3. The results agree well with the theoretical analysis, proving the ability of arbitrary dividing of OAM.

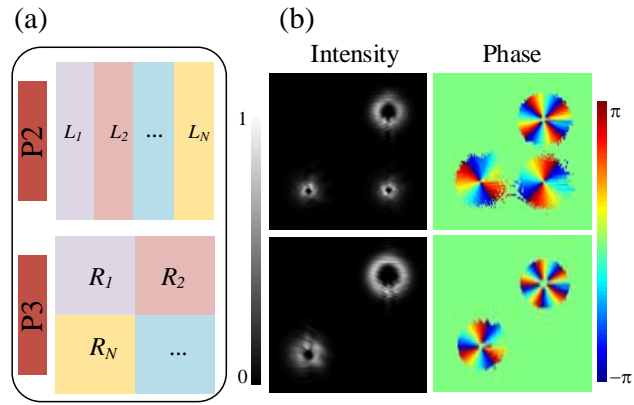

FIG. 7. (Color online) (a) Regional division for arbitrary dividing of OAM in Plane P2 and P3. (b) Simulated results in Plane P3 for arbitrary dividing of OAM when an OAM8 is incident.

## V. CONCLUSIONS

In summary, we report a passive scheme to realize the dividing of OAM in proportion for the first time. The device converts the input light into multiple output light. The OAM of zeroth-order diffracted light is the product of the input OAM and the scaling parameter. The residual light is output from other diffracted orders. Furthermore, we extend the scheme to realize equal *N*-dividing of OAM and arbitrary dividing of OAM. The dividing of OAM shows huge potential for OAM-based classical and quantum information processing.

## ACKNOWLEDGMENTS

The authors acknowledge funding from the National Basic Research Program of China (2011CB301704); Program for New Century Excellent Talents in Ministry of Education of China (NCET-11-0168); Foundation for the Author of National Excellent Doctoral Dissertation of China (201139); National Natural Science Foundation of China (11174096 and 61475052); Foundation for Innovative Research Groups of the Natural Science Foundation of Hubei Province (2014CFA004).

## REFERENCES

[1] D. G. Grier, Nature **424**, 810 (2003).
[2] J. E. Curtis and D. G. Grier, Phys. Rev. Lett. **90**, 133901 (2003).
[3] M. P. Lavery, F. C. Speirits, S. M. Barnett, and M. J. Padgett, Science **341**, 537 (2013).
[4] M. P. J. Lavery, S. M. Barnett, F. C. Speirits, and M. J. Padgett, Optica **1**, 1 (2014).
[5] G. Molina-Terriza, J. P. Torres, and L. Torner, Nature Phys. **3**, 305 (2007).
[6] A. Mair, A. Vaziri, G. Weihs, and A. Zeilinger, Nature **412**, 313 (2001).
[7] N. Bozinovic, Y. Yue, Y. Ren, M. Tur, P. Kristensen, H. Huang, A. E. Willner, and S. Ramachandran, Science **340**, 1545 (2013).
[8] J. Wang *et al.*, Nature Photon. **6**, 488 (2012).
[9] L. Allen, M. Beijersbergen, R. Spreeuw, and J. Woerdman, Phys. Rev. A **45**, 8185 (1992).

[10]     E. Karimi, S. A. Schulz, I. De Leon, H. Qassim, J. Upham, and R. W. Boyd, Light: Sci. Appl. **3**, e167 (2014).
[11]     L. Marrucci, C. Manzo, and D. Paparo, Phys. Rev. Lett. **96**, 163905 (2006).
[12]     E. Rueda, D. Muñetón, J. A. Gómez, and A. Lencina, Opt. Lett. **38**, 3941 (2013).
[13]     M. Q. Mehmood, C.-W. Qiu, A. Danner, and J. Teng, Journal of Molecular and Engineering Materials **02**, 1440013 (2014).
[14]     H. Huang *et al.*, Opt. Lett. **39**, 1689 (2014).
[15]     H. Huang, Y. Yue, Y. Yan, N. Ahmed, Y. Ren, M. Tur, and A. E. Willner, Opt. Lett. **38**, 5142 (2013).
[16]     Y. Yue, H. Huang, N. Ahmed, Y. Yan, Y. Ren, G. Xie, D. Rogawski, M. Tur, and A. E. Willner, Opt. Lett. **38**, 5118 (2013).
[17]     G. C. G. Berkhout, M. P. J. Lavery, J. Courtial, M. W. Beijersbergen, and M. J. Padgett, Phys. Rev. Lett. **105**, 153601 (2010).
[18]     J. Leach, M. J. Padgett, S. M. Barnett, S. Franke-Arnold, and J. Courtial, Phys. Rev. Lett. **88**, 257901 (2002).
[19]     W. Zhang, Q. Qi, J. Zhou, and L. Chen, Phys. Rev. Lett. **112**, 153601 (2014).
[20]     M. Mirhosseini, M. Malik, Z. Shi, and R. W. Boyd, Nat Commun **4** (2013).
[21]     M. P. J. Lavery, D. J. Robertson, G. C. G. Berkhout, G. D. Love, M. J. Padgett, and C. Johannes, Opt. Express **20**, 2110 (2012).
[22]     H. Huang *et al.*, Sci. Rep. **5** (2015).